\documentstyle[aaspp4]{article}

\newcommand{\dd}{{\rm d}}
\newcommand{\pc}{{\rm{pc}}}
\newcommand{\kpc}{{\rm{kpc}}}
\newcommand{\gyr}{{\rm{Gyr}}}
\newcommand{\msun}{{\rm M_\odot}}
\newcommand{\kms}{{\rm{km}\,\rm{s}^{-1}}}
\newcommand{\etal}{{et al}}
\newcommand{\OmegaP}{{\Omega_{\rm P}}}
\newcommand{\csnd}{v_{\rm s}}
\newcommand{\vgrp}{v_{\rm g}}

\newcommand{\rt}{R_{\rm t}}

\newcommand{\eq}[1]{\begin{equation} #1 \end{equation}}
\newcommand{\eqn}[2]{\begin{equation}\label{eq:#1} #2 \end{equation}}
\newcommand{\eqref}[1]{(\ref{eq:#1})}

\begin{document}

\title{Density Waves Inside Inner Lindblad Resonance:\\
Nuclear Spirals in Disk Galaxies}
\author{Peter Englmaier and Isaac Shlosman}
\affil{Department of Physics and Astronomy, University of Kentucky,
  Lexington, KY 40506-0055}
\authoremail{ppe@pa.uky.edu, shlosman@pa.uky.edu}

\begin{abstract}
We analyze formation of grand-design two-arm spiral structure in the
nuclear regions of disk galaxies. Such morphology has been recently
detected in a number of objects using high-resolution near-infrared
observations. Motivated by the observed (1) continuity between the
nuclear and kpc-scale spiral structures, and by (2) low
arm-interarm contrast, we apply the density wave theory to explain the
basic properties of the spiral nuclear morphology. In particular, we
address the mechanism for the formation, maintenance and the 
detailed shape of nuclear spirals. 
We find, that the latter depends mostly on the shape of the underlying
gravitational potential and the sound speed in the gas. Detection of
nuclear spiral arms provides diagnostics of mass distribution within the
central kpc of disk galaxies. Our results are supported by 2D numerical
simulations of gas response to the background
gravitational potential of a barred stellar disk. We investigate the
parameter space allowed for the formation of nuclear spirals using a
new method for constructing a gravitational potential in a barred
galaxy, where positions of resonances are prescribed.
\end{abstract}


\section{Introduction}

Recent high-resolution near-infrared (NIR) observations of central 
regions in disk galaxies have revealed nuclear spirals on scales of
a few 100~pc. Typically these spirals are small-scale flocculent 
patterns, as in NGC~278 
(Phillips \etal. 1996), M51 (Grillmair \etal. 1997), M81 (Devereux, Ford 
\& Jacoby 1997), M87 (Ford \etal. 1994; Dopita \etal. 1997), NGC~2207 
(Elmegreen \etal. 1998) and others (Carollo, Stiavelli \& Mack 1998). 
In a few cases, the grand-design nuclear spirals that connect to the outer 
spiral structure were observed (Laine \etal. 1999; Regan \& Mulchaey 
1999). 

The arm-interarm contrast in the region of nuclear spirals is low,
typically less than 0.1 mag. This may be evidence of a low-amplitude
density perturbation in the nuclear region, which is not associated with
star formation. The nuclear spirals are best observed in NIR color
index images,
most prominently in $J$-$K$. Red colors of nuclear spirals
suggest the importance of dust and hence of the gas. Nuclear structure 
is sometimes encircled by often incomplete rings of star formation (like 
in NGC~5248). 
Such rings typically form in the neighborhood of inner Lindblad resonance(s)
(ILRs) (e.g., Buta \& Combes 1996), and indicate an ongoing radial
gas inflow in the disk towards the ring, due to non-axisymmetric perturbations
in the background gravitational potential.  
So far, the majority of all grand-design nuclear spirals involve disks with 
weak to moderately weak\footnote{%
as measured by deprojected bar axial ratios.} 
bars, in some cases detected only in high-resolution NIR imaging 
(Knapen, Shlosman \& Peletier 1999). The frequency of occurrence of
nuclear structures in disk galaxies is unknown. 

Density wave theory supported by observational evidence predicts that
the kpc-scale {\it stellar} spiral structure should end well outside
the region where nuclear spirals are found (e.g., Toomre 1969;
Goldreich \& Tremaine 1979; Binney \& Tremaine 1987). The reasons for
this are discussed in Section~2. Instead, {\it gas} density waves can
generate spiral structure at all radii, but have not been considered
so far.  Numerical studies of the stationary gas response to an
imposed bar-like perturbation have been performed by many authors
(e.g., van Albada \& Roberts 1981; Schwarz 1981; van Albada \& Sanders
1982; Mulder 1986; Athanassoula 1992). These studies have focused
on the large scale structure of barred galaxies with insufficient
resolution in the central regions.  It is the continuity of the
grand-design spiral structure from about 100~pc to a few kpc, and the
low amplitude they have in the center, that stimulated our interest in
gas-dominated density waves. Dust is expected to have a similar
distribution.

In this paper, we focus on generating and maintaining the spiral
morphology in the gas deep inside the ILR. We use a large-scale
stellar bar in order to drive the spiral shocks across the ILR and
further inwards.  We find that for a particular range in central mass
distribution and sound speeds, steady-state low-amplitude spiral
pattern can be maintained indefinitely. Our view is supported by 2D
numerical simulations of gas response inside the corotation radius.

This paper is structured as follows. Section~2 is devoted to relevant
aspects of the density wave theory. Section~3 describes the numerical
model used to evolve the gaseous component in disk galaxies.
Section~4 presents our results and compares them with theory,
and Section~5 is devoted to discussion and conclusions.

\section{Theory of spiral waves}

We introduce briefly some important concepts from the density wave
theory developed by Lin \& Shu (1964) and others. For an in-depth 
discussion we refer the reader to Toomre (1977; 1981), Binney \& 
Tremaine (1987), and Bertin
\etal. (1989). A self-gravitating stellar disk with a surface density
$\Sigma$, velocity dispersion $\sigma_*$ and epicyclic frequency
$\kappa$, supports two wave modes: short and long. Stellar spiral arms
can be made out of wave packets of either type of wave. These waves
can only propagate in certain radial domains separated by resonances,
while outside these domains the wave trains decay. 

While advected 
azimuthally with a fraction of a fluid velocity,
the waves move radially with a group
velocity $\vgrp={\partial\omega(R,k)/\partial k}$, where $\omega$ is
the wave frequency in the disk at radius $R$, and $k$ is a wave vector
(Toomre 1969; Whitham 1974). For Toomre parameter $Q\equiv \sigma_*
\kappa/\pi G \Sigma > 1$, the corotation resonance is surrounded by a
forbidden region where the stellar density waves are evanescent, i.e.,
waves decay. At the boundary of this region, short waves are reflected
into long waves and vice versa. At the inner and outer Lindblad 
resonances\footnote{The ILRs and OLRs are resonances between the precession  
frequencies $\Omega\pm\kappa/2$ and the pattern speed $\OmegaP$.}
(ILRs, OLRs), the long waves are reflected (but change from 
leading to trailing waves and vice versa) and the short ones are 
absorbed. The stellar waves of both types can
not propagate across the ILRs and OLRs. Moreover, stellar waves
cannot propagate inside the ILRs even if they are generated
there! Goldreich \& Tremaine (1978; 1979) studied propagation
of gaseous density waves and found that they behave in a similar
fashion except at the Lindblad resonances, where short gaseous waves 
are able to propagate through the ILR all the way to the center (or through
the OLR to infinity).  In this paper, we argue that such waves
can create a stationary spiral pattern in the nuclear disk inside the
ILR in some potentials. We specifically target the case where the 
large-scale perturbation is caused by the stellar bar with 
a pattern speed $\OmegaP$ and therefore, assume $m = 2$.
However, this can be generalized for any symmetry of the perturbation.
Detection of such nuclear spirals hence
provide diagnostics of the mass distribution within the central few
hundred pc.

Without self-gravity, i.e., $Q\rightarrow \infty$, stellar waves
cannot propagate in the disk at all, because $\omega$ is
independent of $k$ and $v_g=0$. However, nonwave-forced response 
is expected in non-axisymmetric potentials (e.g., Goldreich \& Tremaine 
1979). Non-self-gravitating gaseous disks
are able to support waves because of the pressure forces, but
only inside the ILR (and outside the OLR), and under
conditions outlined below.  To show this explicitly, we write the
linearized dispersion relation for mode $m=2$ in the gaseous disks
without self-gravity and embedded in an external frozen potential:
\eqn{disp}{
   (2\Omega-\omega)^2 = \kappa^2 + k^2\csnd^2,
}
where $\csnd$ is the sound speed. The applicability of this equation
in principle is limited by the condition $kR \gg 1$, or in other
words, to wavelengths short compared with $R$ (e.g., Binney \&
Tremaine 1987). Close to the ILR, $k\to0$ and hence the
WKB (Wentzel-Kramers-Brillouin) approximation fails here. Eq.~\eqref{disp}
can be re-written in the form
\eqn{eq2}{
\left({k\csnd\over 2}\right)^2 = (\Omega-\kappa/2-\OmegaP)(\Omega+\kappa/2-\OmegaP),
}
where the spiral pattern speed $\OmegaP=\omega/2$. Inside the
corotation radius, the gaseous waves can propagate only when
$\Omega-\kappa/2-\OmegaP > 0$. When one ILR is present, they
exist between the center and the ILR. When two ILRs are
present, they exist between those resonances.
Using the definition of the wave group velocity (Toomre 1969)
and Eq.~\eqref{eq2}, we obtain
\eqn{vgrp}{
   \vgrp= {k \csnd^2\over\omega-2\Omega}
        = \csnd{\sqrt{(\Omega-\kappa/2-\OmegaP)(\Omega+\kappa/2-\OmegaP)}
	\over \OmegaP-\Omega}.
}
For $\Omega-\kappa/2\gg\OmegaP$, i.e., close to the center, the group
velocity is close to the sound speed. It is precisely where the nuclear
arms cannot exist due to a large shear.

   It is important to understand that the above treatment of wave
propagation is limited to the regions where no shocks exist and the
wave amplitude can be considered linear. When a bar perturbation is
present, the vicinity of the ILR (or OLR) is expected to host
shocks, due to the response of the near-circular orbits to
the bar torquing (e.g., Sanders \& Huntley 1976). This response is 
aligned with the bar
between the corotation and the ILR, and is $\pi/2$ out of phase with
the bar, inside the ILR (when only one ILR is present).
Orbits aligned with the bar are called $x_1$ orbits in the notation of
Contopoulos \& Papayannopoulos (1980).  The change of the response
across the resonance will cause the flow streamlines to intersect,
forming shocks, encompassing about $\pi$ 
radians, as a result. Deeper inside the ILR, the gas settles on
non-intersecting orbits elongated perpendicular to the bar major axis
and hence no shocks are expected to be present all the way to the
center as these orbits co-rotate with the bar. (These orbits oriented 
perpendicular to the bar's major axis are called $x_2$ orbits.) It is 
this inner region which is of interest to us and where the nuclear 
spirals have been observed.

The shape of a nuclear spiral can be inferred from the density wave
theory. Material arms in disk galaxies can not survive
for many rotations because of the differential rotation in the
disk. Any material wave will be quickly sheared and become tightly
wound with the pitch angle $i$ tending to $0$, and no steady state can
be achieved. A radially propagating density wave, in general, will be
winding as well, albeit slower than a material wave. However, if it 
is continuously excited at some $R$ at a fixed phase angle,
e.g., with respect to a large-scale bar rotating with a pattern speed
$\OmegaP$, the resulting spiral shape will be stationary in the bar's
frame. 

The spiral shape is given by the ratio of the wave's radial
to azimuthal phase velocities (Binney \& Tremaine 1987), i.e. in
the frame co-rotating with the bar, the pitch angle of the spiral is
\eqn{incl}{\tan i = \left|{ 2\over kR }\right|,}
where $m=2$ was assumed. The azimuthal phase velocity of the spiral
in the bar presence is $\OmegaP$. Integrating the geometrical 
identity $\tan i = \dd R/R \dd\varphi$,
yields the shape of the spiral pattern, i.e., the azimuthal
angle $\varphi(R)$:
\eq{
\varphi(R) = \int_R^{R_0}{{ k(\bar R) \over m}
	{\dd \bar R}} + \varphi_0,
}
where $\bar R$ is a dummy variable. As we show below for certain
potentials, the shape of the resulting spiral pattern corresponds to
the observed one and can be maintained, for example, by the gas
response to the stellar bar.

\section{Gas Dynamical Simulations}

In order to test the pitch angle of the nuclear spiral pattern, given by
Eq.~\eqref{incl}, we resort to 2D hydrodynamical simulations of gas
response to non-axisymmetric perturbations in disk galaxies. We first
describe the construction of the background stellar potential and give
details of initial setup and numerical method. Results from the simulations
are analyzed in Section~\ref{sec-modresults}.

\subsection{Constructing the stellar potential: the standard model}
\label{sec-model}

The linear theory provides the locations of resonances between the bar
pattern speed and the orbital precession rate
\eqn{res}{  
	f(R)\equiv\Omega(R)-\kappa(R)/2.
}
In order to study the influence of ILRs on the gas flow in barred
galaxies, we constructed a model potential which prescribes the
locations of resonances (Englmaier \& Shlosman 1999). For the ILRs,
this is achieved by specifying the function $f(R)$, corresponding to a
particular mass distribution inside the corotation. Outside the
corotation we use the analytical potential described in the Appendix.
As found in Section~4, a nuclear spiral forms when only a single ILR
is present, and therefore we limit our discussion to this case.
Multiple ILRs will add additional spiral structure on yet smaller
scales complicating the observed picture.

\placetable{tbl-1}

\begin{figure}
\epsscale{.4}
\plotone{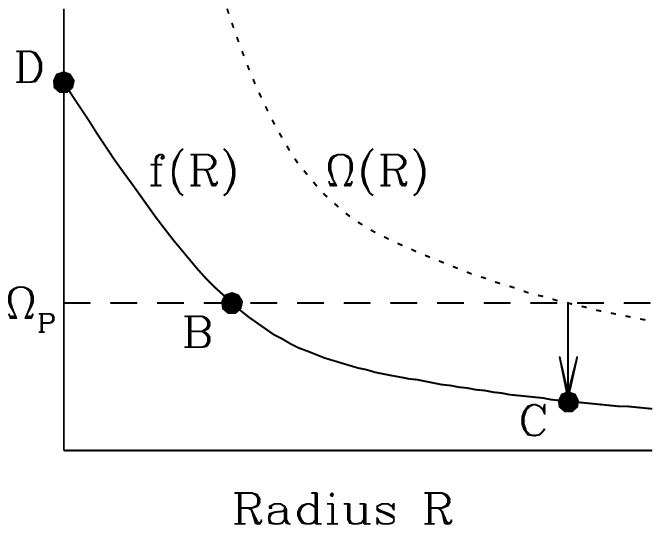}
\caption{Model construction. Points D, B and C correspond to the
$f_0\OmegaP$, ILR and corotation, respectively. $\OmegaP$
represents the bar pattern speed, $f(R)=\Omega-\kappa/2$ is
the orbital precession frequency, and $f_0 \equiv f(R=0)/\OmegaP$ (see text 
for further explanations).
\label{fig-toy}}
\end{figure}

We chose cubic splines for $f(R)$ representation
(Fig.~\ref{fig-toy}). The method described below is general.  The spline
is divided in two cubic spline segments between points D, B, and C
(see Fig.~\ref{fig-toy}). The spline parameters are summarized in
Table~1. Parameter $C_3$ is chosen so that the first
derivative $f'(R)$ is continuous at point B.  The spline between
B and C is fitted to an analytical potential at corotation
(C). The position of the OLR is given by the outer potential, although
in principle we could extend our method to control the position of
this resonance as well. Our choice of $f(R)$ corresponds to a
physically meaningful mass distribution and rotational velocity curve
(see below). The pattern speed and fitting parameters $E_1$ and $E_2$
(Table~1) are fixed by the position of the corotation radius. For all
the models presented here, the corotation is at $6\,\kpc$ and the single
ILR is at 2~kpc, corresponding to the pattern speed of
$24.9\,\gyr^{-1}$, or one rotation in $0.25\,\gyr$.  

\placefigure{fig-toy}

\begin{figure}
\epsscale{.4}
\plottwo{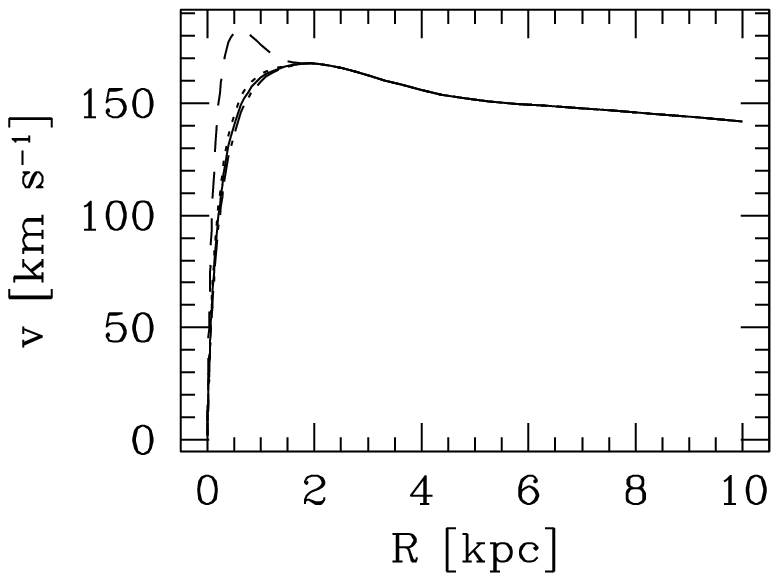}{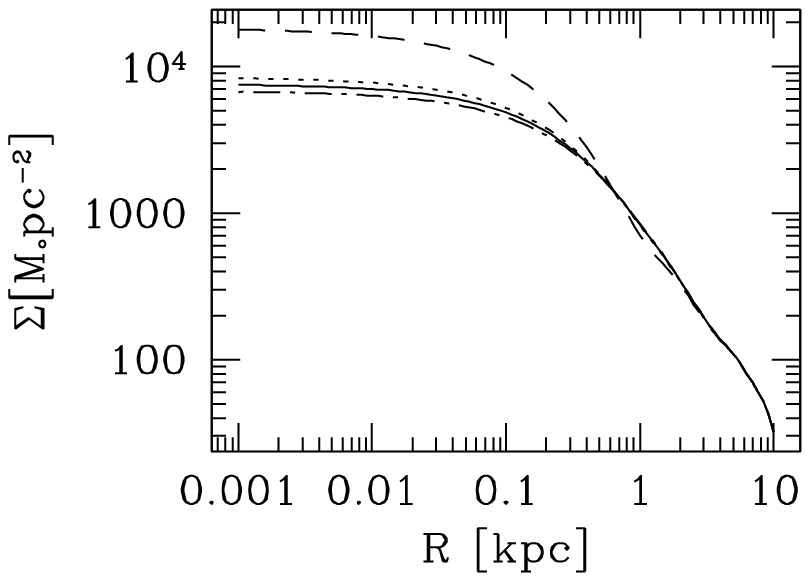}
\caption{Rotation curve (left) and surface density (right) for all 
stellar disk models. Models shown are the standard model with $f_0=3$ 
(solid), $f_0=2.7$ (dash-dotted), $f_0=3.3$ (dotted), and $f_0=6$ (dashed).
\label{fig-surf}}
\end{figure}

Next, we calculate the model potential corresponding to the resonance
curve $f$.  Using the definition of the epicycle frequency $\kappa(R)$
and Eq.~\eqref{res}, we obtain an ordinary differential equation for
$\Omega(R)$:
\eq{
   \Omega'(R) = 2 f(R) {f(R) - 2\Omega(R) \over R \Omega(R)}.
}
This ODE is the 2nd kind Abel-type equation, which has no analytical
solution for any interesting choice of $f(R)$. We solve it numerically
with the Runge-Kutta method. Using the numerical solution for
$\Omega(R)$, we compute the potential in the galactic plane
\eq{
\Phi(R)=\int_\infty^R{\bar R \Omega^2(\bar R)\dd \bar R}
}
and the rotation curve $v(R)=R \Omega(R)$. Figure~\ref{fig-surf}~(left)
shows that the rotation curve is raising steeply within the first kpc
and stays approximately constant thereafter.

\placefigure{fig-surf}

A physically meaningful potential in the plane must correspond to a
positive surface mass density everywhere, which is given in
Fig.~\ref{fig-surf}~(right) for our model, using a prescription by 
Binney \& Tremaine (1987), namely
\eqn{surden}{
\Sigma(R)={1\over G\pi^2}\left[
{1\over R}\int_0^R{{\dd v^2\over\dd\bar R}K(\bar R/R)\dd\bar R
+\int_R^\infty{{1\over\bar R}{\dd v^2\over\dd\bar R}K(R/\bar R)\dd\bar R}
}\right].
}
Figure~\ref{fig-surf}~(right) shows that for the parameter space under
consideration, $\Sigma(R)>0$ everywhere in the model.

Finally, we add the Ferrers' bar potential (Ferrers 1877) to the
axisymmetric model. The density in the bar is given by
\eq{
    \rho_{\rm b}(R) = \rho_0 \left(1 - {x^2\over a^2}-{y^2+z^2\over b^2}\right),
}
where the semi-major axis $a=5\,\kpc=R_{\rm cr}/1.2$, the semi-minor
axis $b=2\,\kpc$, $\rho_0={15 M_{\rm B}/(8\pi a b^2)}$ is the central
density, and $M_{\rm b}=5\times10^9\,\msun$ is the mass of the
bar. Here $R_{\rm cr}$ is the corotation radius. We assume a massless
Ferrers' bar and add only the non-axisymmetric part of the Ferrers'
potential to the disk. The Eq.~\eqref{surden} was used again to check
for a positive density everywhere in the final model.  The advantage
of this method is that the linear resonances remain in the prescribed
positions.

We refer to the standard model as the one with $f_0\equiv
f(R=0)/\OmegaP = 3$, and the sound speed $\csnd = 10\,\kms$. In
addition, a number of models are used with different values of $f_0$
and $\csnd$ (see Section~\ref{sec-checkeq4}).

\subsection{Solving the hydrodynamical equations}
\label{sec-hydro}

The continuity and Euler equations are solved in 2-D using ZEUS-2D
from Stone \& Norman (1992a, 1992b).  To maximize the resolution in the
center, we use a polar grid with 124 logarithmically equidistant
radial grid points between $5\,\pc$ and $10\,\kpc$ and 100 azimuthal
grid points between $0$ and $\pi$.  We assume a point symmetry with
respect to the center and, therefore, impose a periodic boundary
condition at the azimuthal boundaries. Test calculations without this
symmetry constraint did show the same result with differences at the
level of grid resolution. At the inner boundary of the radial grid, we
implement an outflow condition, that is, the gas can only leave the
grid. To stabilize the inner boundary against the spurious wave
reflections, we average the velocity field in the cells adjacent to
the boundary. At the outer boundary, we set the radial velocity to
zero, and assume a constant azimuthal velocity and density across the
boundary.

We assume an isothermal equation of state for the gas which is
believed to provide a good description of the large-scale single-phase
ISM behavior (e.g., Cowie 1980). The initial gas setup is that of an
exponential disk
$\Sigma(t=0)\equiv\Sigma_0(R)\propto\exp(-R/2\,\kpc)$.  The gas is
non-self-gravitating and resides initially on circular orbits.  The
bar is gradually turned on within one rotation, allowing the gas to
settle in its preferred flow configuration. In order to reach a steady
state, and to avoid the usual problem of depopulating the disk of its gas
content, we include the `gas recycling' $\dd\Sigma/\dd
t=0.3\,\gyr^{-1}\Sigma_0[1-(\Sigma/\Sigma_0)^2]$.  The latter mimics
the gas loss at the density peaks due to the star formation, and
compensates for the gas loss due to the inflow towards the center in
other places (e.g., Athanassoula 1992).

In order to see whether our results depend on the numerical
resolution, we repeat the standard model with three times higher
radial resolution. The spiral appears at almost exactly the same
location. The small deviation comes from superior resolution of the
shock front. 

\section{Results of modeling and comparison with theory}
\label{sec-modresults}

\begin{figure}
\epsscale{1.0}
\plotone{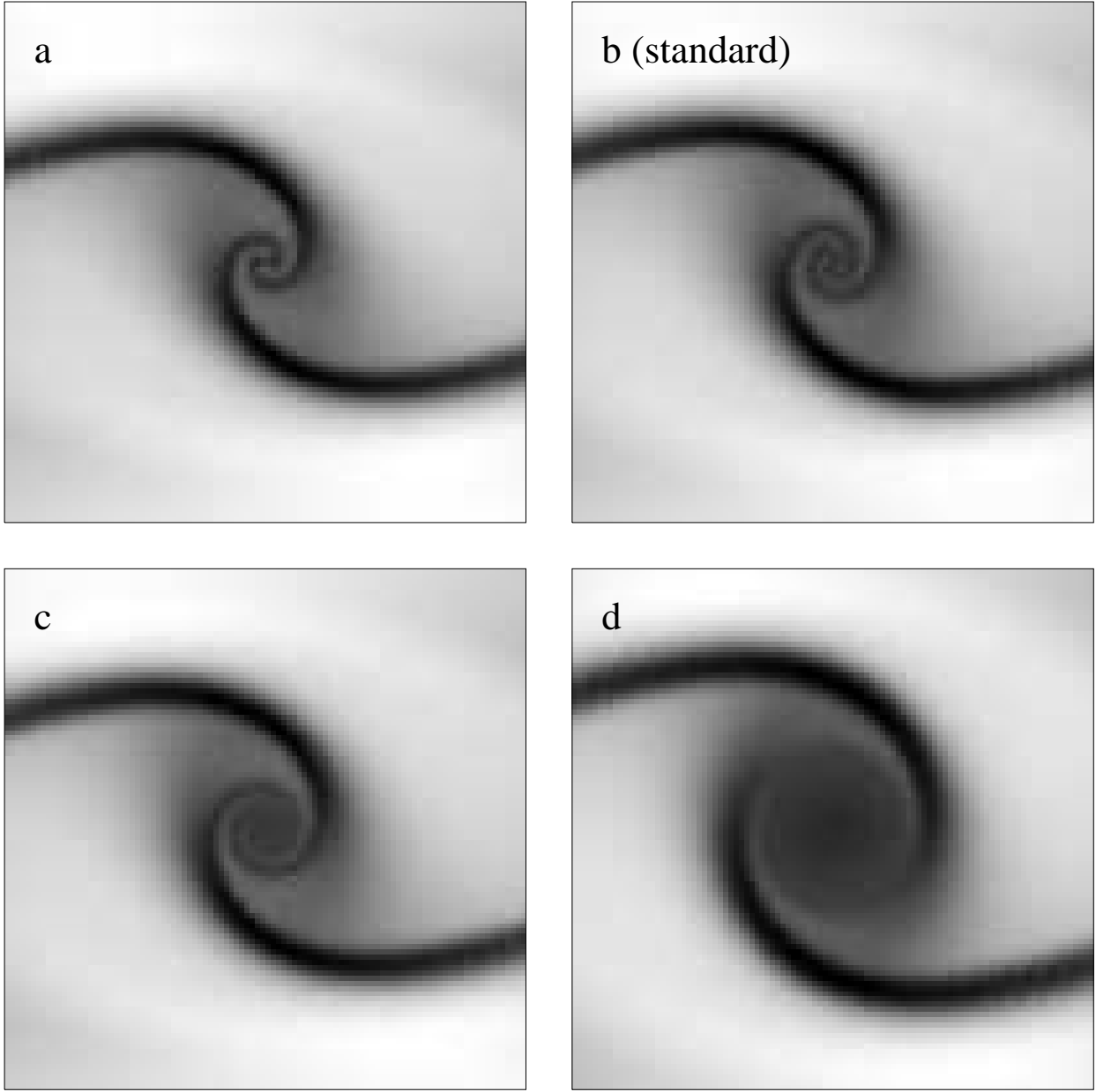}
\caption{Grey-scale images of steady state gas response to the stellar
bar torquing. Only the inner $2\,\kpc$ are shown out of the
model extending to $10\,$kpc. The stellar bar is horizontal and extends
to $5\,$kpc; the gas rotation is clockwise.
Individual frames differ by the value of $f_0$: (a) $f_0=2.7$, (b) 
$f_0=3$, (c) $f_0=3.3$, and (d) $f_0=6$. 
The sound speed is $\csnd=10\,\kms$ and the single ILR is at $2\,\kpc$.  
\label{fig-rho-A}}
\end{figure}

For a certain parameter range in $f_0$ and $\csnd$, our model
generates a spiral pattern between the corotation and the center,
winding an unusual $\sim 3\pi$ radians angle (e.g.,
Fig.~\ref{fig-rho-A}b). The part of the pattern corresponding to the
winding of $\pi$ radians, i.e., between the corotation and the first crossing
of the bar major axis, is a strong standing shock in the bar frame. It
can be understood in terms of the gas response to the bar forcing. We
refer to the radius when the shock first crosses the bar major axis as a
transition radius, $\rt$.  The spiral pattern is formed by the gas
moving from the $x_1$ orbits aligned with the bar to the $x_2$ orbits
perpendicular to the bar (see Section~2). This was
confirmed by means of nonlinear orbit analysis (described in Heller \&
Shlosman 1996). Shocks following the spiral pattern along the leading
edges of the bar are observed as dust lanes in real galaxies and
reproduced in numerical simulations (Prendergast 1962; Athanassoula
1992).  The modeled gas response in Fig.~\ref{fig-rho-A}b, however,
clearly shows the continuation of the spiral pattern well past the region
of expected winding.

\placefigure{fig-rho-A}

\begin{figure}
\epsscale{.5}
\plotone{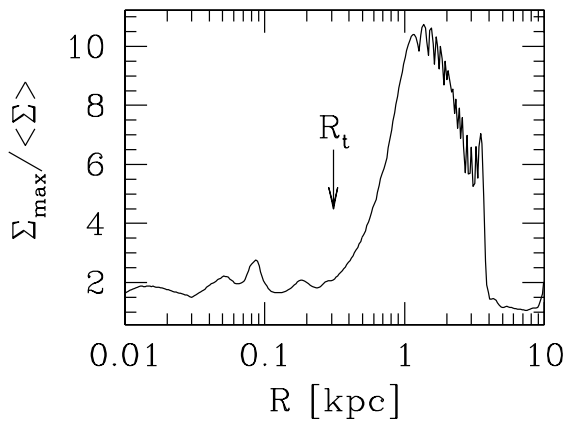}
\caption{Spiral arm surface density normalized by the average density at
radius $R$ for the standard model with $3\times$ radial resolution
of Fig.~\ref{fig-rho-A}b. The spiral weakens dramatically inside $\rt$ and
outside the $4\,$kpc where its crosses the bar axis again. Shocks in 
the outer spiral ($>\rt$) are unresolved, thus the surface density plotted 
here is only a lower limit.
\label{fig-jump}
}
\end{figure}

To understand the reason for the observed gas response and the large
winding angle, we compare the density response amplitudes along the
spirals (Fig.~\ref{fig-jump}). We note first that this amplitude
decreases substantially after the first $\pi$, counting from the
corotation inwards, i.e., inside the $\rt$. Such decline means that
the shock strength drops sharply and the gas response at smaller
radii is more of a linear wave.  This is consistent with our
understanding that the gas settles on the $x_2$ orbits inside the ILR,
and the initial $\pi$ winding corresponds to this process.  At smaller
radii, another process must support the modeled spiral pattern, a
process, which requires a milder orbital change. In Section~2 we have
discussed this process within the framework of the density wave
theory.

\placefigure{fig-jump}

Secondly, we observe that the spiral pattern is stationary in the bar
frame. Viewed in the fluid rest frame at some $R$, the radial
component of the wave vector points inwards, because the gas rotates
faster than the spiral pattern everywhere inside the corotation. It is 
convenient to explain the spiral in terms of a
wave packet propagating radially inwards, which is sheared by the
differential rotation in the disk.  The only waves which can propagate
inside the ILR are the short waves, and their group velocity is slower
than the sound speed (Eq.~\ref{eq:vgrp}), due to the contribution of
the epicyclic motion.

\begin{figure}
\epsscale{1.0}
\plotone{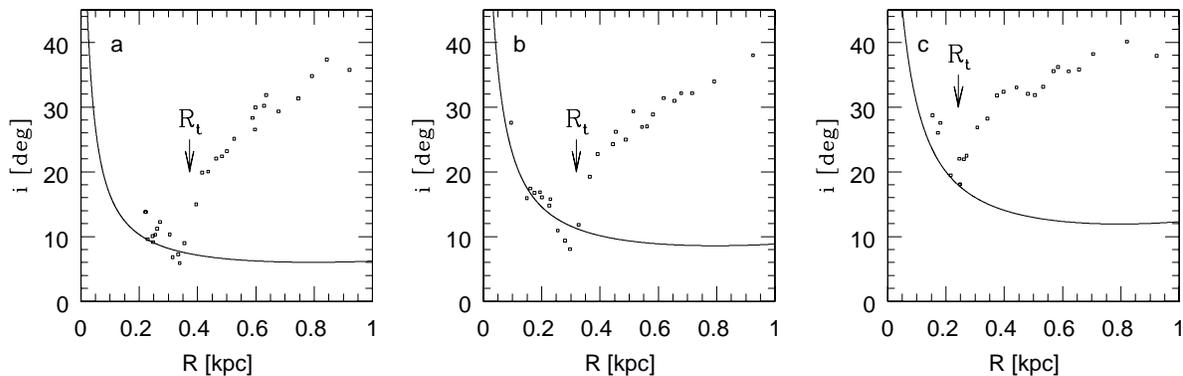}
\caption{Spiral pitch angle vs. $R$, for models with $f_0=2.7,\,3,\,3.3$ 
(from left to right), and sound speed $\csnd=10\,\kms$. Theoretical 
(solid line) and modeled (dots) curves are shown. The vertical arrow 
shows the position of transition radius $\rt$ where the spiral pattern 
winds up the initial angle of $\pi$ (measured from corotation).
\label{fig-incl-A}
}
\end{figure}

The low arm-interarm contrast observed in the model inside $\rt$
(Fig.~\ref{fig-jump}) suggests that the wave is not accompanied by a
strong compression. This is in agreement with the observed nuclear
spirals which do not exhibit extensive star formation and are mainly
seen because of the dust obscuration. This partly supports our neglect
of the self-gravitational effects in the gas and the use of the
formalism developed in Section~2. Equation~\eqref{incl} estimates the
pitch angle of the gas response under these circumstances. Most
importantly, the pitch angle of the innermost spiral pattern decreases
quickly with the radius, i.e., the pattern becomes tightly wound. This
is in sharp contrast with the pitch angle at $R > \rt$, which
corresponds to the $x_1$-to-$x_2$ orbital switch. Here the pitch angle
quickly increases with radius, and the overall response becomes
open. The wedge-shaped profile of the pitch angle in
Fig.~\ref{fig-incl-A} is therefore characteristic of switching from
one type of response to another.

\placefigure{fig-incl-A}

Both responses cannot coexist at the same radii.  Moving
inwards from the ILR, the spiral pattern associated with the outer
shocks winds around the center and the pitch angle decreases. At some
value, $\rt$, the low-amplitude density waves take over and slowly
increase this angle. The gas can only sustain waves of a
particular frequency (and wave vector) at each $R$, those are
solutions to the dispersion relationship (Eq.~\ref{eq:disp}). 
Therefore, the solutions for the pitch angle can only
switch from the inner to the outer one where they are comparable.

\label{sec-checkeq4}

In order to test Eq. \eqref{incl}, we run two series of models. The
first series varied $f_0$, thus changing the slope of the resonance
curve $f(R)$ between the center and the ILR, keeping all other
parameters of the model constant.  The second series changed the sound
speed in the gas, keeping $f_0$ constant.  The pitch angle $i$ of the
modeled spiral was measured by assuming its local logarithmic
shape. This is equivalent to measuring the slope of the spiral in the
$(R,\varphi)$ coordinate plane by fitting a tangent to it.  In each
model, we measured the pitch angle at different radii to probe the
wave propagation.

\begin{figure}
\epsscale{1.0}
\plotone{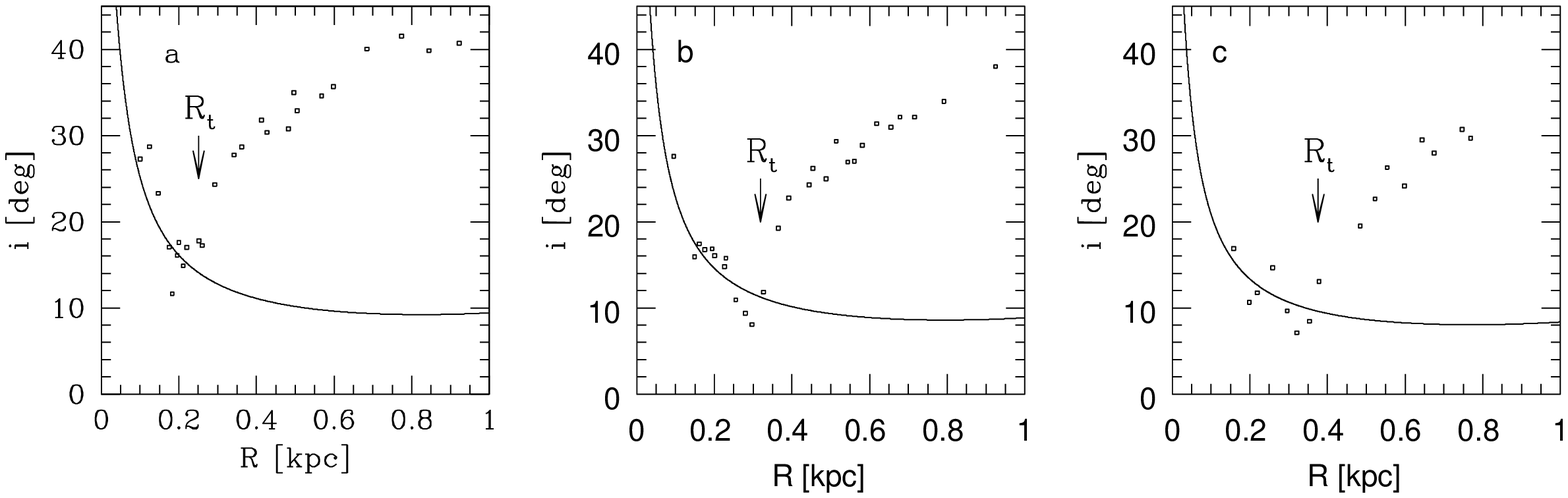}
\caption{Spiral pitch angle vs. $R$, for models with $f_0=3$ and
the sound speed $\csnd=7,\,10,\,14\,\kms$
(from left to right). Theoretical (solid line) and 
modeled (dots) curves are shown. The vertical arrow shows the position
of transition radius $\rt$ where the spiral pattern winds up the initial
angle of $\pi$ (measured from corotation). 
\label{fig-incl-c}
}
\end{figure}

\placefigure{fig-incl-c}

Fig.~\ref{fig-incl-A} reveals a strong dependence of the spiral shape
in the nuclear region on $f_0$. Most important, the winding is
increased with $f_0$. A factor of 2 in $f_0$ makes the spiral pattern
so tightly wound that it is not recognizable any more beyond the
initial winding of $\pi$. The result is the formation of a featureless
nuclear disk within the inner $400-500$~pc. For larger disks,
corresponding to larger $f_0$, the incoming shocks from the leading
edges of the bar move away from the bar major axis.  For decreasing
$f_0$, the spiral becomes less tightly wound and finally collapses
towards the bar's major axis. The reason for this behavior is that
decreasing $f_0$ corresponds to decrease in the central mass
concentration in the model and in the strength of the ILR. The modeled
gas responds in a nonlinear fashion, staying on $x_1$ orbits inside
the weak ILR. This effect was observed in numerical simulations
(Englmaier \& Shlosman 1999) and leads to the centered shocks on the
major bar axis. The strength of this effect depends also on the sound
speed in the gas (Englmaier \& Gerhard 1997), and is discussed below.

Equation~\eqref{incl} predicts the shape of the spiral pattern in the
non-self-gravitating gas (solid line in Fig.~\ref{fig-incl-A}).  The
measured pitch angle $i$ of the modeled spiral is given by dots on the
same figure.  We observe a two-fold behavior for $i$, i.e., from the
ILR inwards $i$ is decreasing until the shock pattern winds the angle
of $\pi$ ($\rt$, indicated by an arrow in Fig.~\ref{fig-incl-A}), and
then starts to unwind.  Remarkably, there is a sharp break in $i(R)$
slope at this point. The inner part of $i(R)$ curve is in agreement
with Eq.~\eqref{incl}. As we have noted above, the
outer part of the $i(R)$ curve describes the nonlinear gas response to
the bar forcing. Fig.~\ref{fig-incl-A} demonstrates that the
large-scale shock penetrates inside the ILR and perturbs the nuclear
gas disk, launching the wave packets towards the center. Hence it is
the large-scale shock which directly drives the nuclear spiral
structure.

\begin{figure}
\epsscale{.35}
\plotone{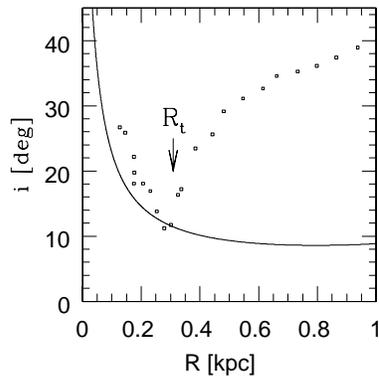}
\caption{
Same as in Fig.~\ref{fig-incl-A}, but with $3\times $ radial grid 
resolution, for standard model with $\csnd=10\,\kms$, and $f_0=3$.
\label{fig-incl-hires}
}
\end{figure}

It is interesting that the measured pitch angle $i$ closely follows
the theoretical $i(R)$ curve within
$\rt$ (Fig.~\ref{fig-incl-A}). The latter was calculated based on the WKB
approximation which requires $kR\gg 1$,
i.e., a tightly wound spiral. However, we observe that even for points
at $i\sim 30^\circ$ the correspondence is very good, and becomes
better when resolution is increased (Fig.~\ref{fig-incl-hires}).

\placefigure{fig-incl-hires}

\begin{figure}
\epsscale{.5}
\plotone{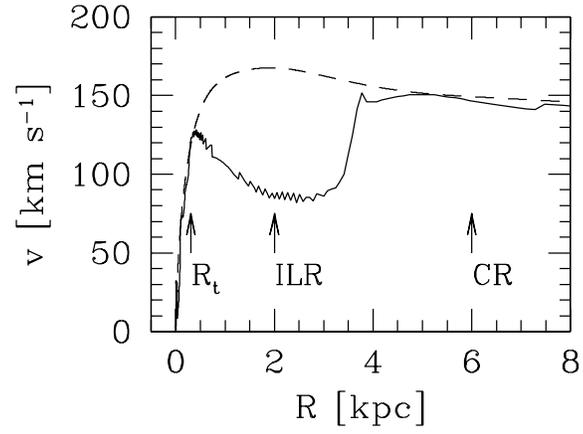}
\caption{The gas azimuthal velocity component (solid line) measured 
along the spiral (i.e., along the density maximum at each $R$), in 
the inertial frame for the standard model. Circular velocity for the 
axisymmetric part of the potential (dashed line) is shown for comparison. 
The gas is on nearly circular orbits inside $\rt$ and outside $4\,\kpc$,
and transits from $x_1$ to $x_2$ orbits in between.
\label{fig-vt}
}
\end{figure}

In Fig.~\ref{fig-vt} we show the azimuthal velocity component of the
gas flow measured along the spiral arm. Inside the transition radius
$\rt$ the azimuthal gas velocity is very close to the rotational
velocity in the axisymmetric potential (dashed line). Somewhat further
out, the tangential velocity shows a sharp drop, due to the
non-circular motions in the gas which is in the process of settling on
$x_2$ orbits. This figure demonstrates the observational significance
of the transition radius, $R_t$. Namely, it represents the outer
radius of the nuclear disk where the gas is found on nearly circular
$x_2$ orbits aligned with the minor axis of the bar. The deviation
from the axisymmetric rotation curve is largest around the ILR
radius. Furthermore, the bar-driven shocks reach the bar major axis
again at $\sim4\,\kpc$. Beyond this radius, the gas follows
nearly circular orbits.

\placefigure{fig-vt}

Note, that with increasing $f_0$, the nuclear spiral becomes
progressively more tightly wound. Because of the finite resolution in
the numerical code, we need up to 4 grid points to resolve the
shock. This corresponds to minimum spiral inclination of about
$7^\circ$, below which the neighboring shocks remain unresolved.  In
such a case, local numerical viscosity is increased and the spiral
waves are quickly damped out. This explains why Fig.~\ref{fig-rho-A}
does not show a nuclear spiral propagating towards smaller $R$. A
similar effect can happen in a realistic disk, where the `shock' width
is given by the mean-free path of individual clouds in the clumpy
ISM. Hence, there is physical significance to the transition
radius, $\rt$, where $i(R)$ is reaching its minimum. Namely, a pitch
angle that is too small at $\rt$ prevents the wave packet from
propagating inwards without interaction with other packets.
The transition radius $\rt$, where the nuclear spiral starts
propagating inwards, depends on $f_0$ and $\csnd$.  For larger $f_0$
and smaller $\csnd$, the transition radius moves out, leaving room for
a larger nuclear disk.

\begin{figure}
\epsscale{.4}
\epsscale{1.0}
\plotone{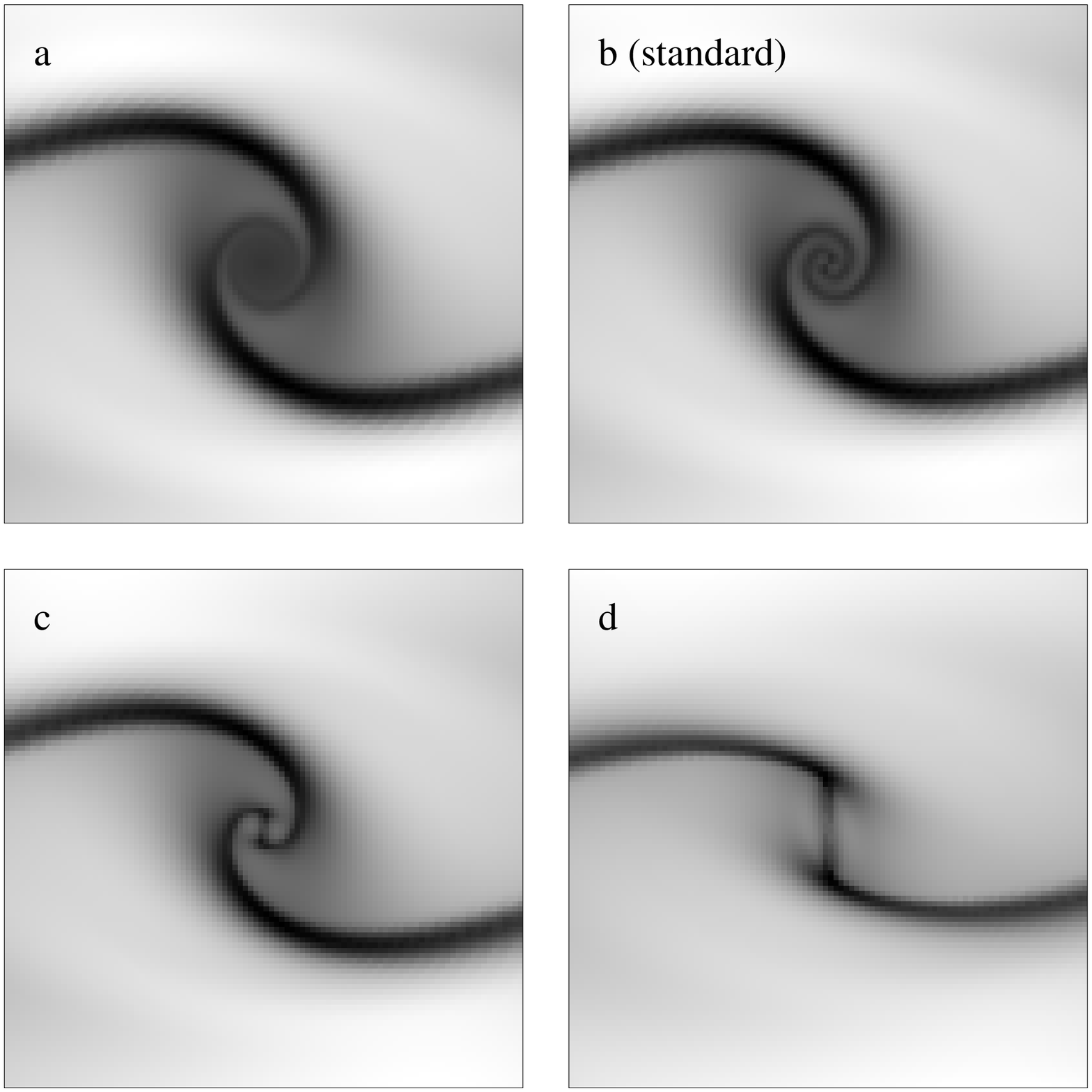}
\caption{Grey-scale images of steady state gas response to the stellar
bar torquing. Only the inner $2\,\kpc$ are shown out of the
model extending to $10\,$kpc. The stellar bar is horizontal and extends
to $5\,$kpc; the gas rotation is clockwise. Individual frames differ by 
the value of sound speed: (a) $\csnd=7\,\kms$, (b) 
$\csnd=10\,\kms$, (c) $\csnd=14\,\kms$, and (d) $\csnd=20\,\kms$.
The $f_0=3$ and the single ILR is at $2\,\kpc$.
\label{fig-rho-csnd}}
\end{figure}

\placefigure{fig-rho-csnd}

Varying the sound speed in the gas is another way of changing the
strength of resonances, the ILR in this case.  With increasing sound
speed, the pitch angle becomes larger and the spiral pattern opens,
until the shocks move closer to the bar major axis
(Fig.~\ref{fig-rho-csnd}).  Eq.~\eqref{incl} predicts this trend due
to the dependence of the wave number $k$ on the sound speed given by 
Eq.~\eqref{eq2}. On
the large scale, this corresponds to shocks being only slightly offset
from the bar major axis, as for low values of $f_0$.  Athanassoula
(1992) observed this gas response of centered shocks when no ILR(s)
were present in the model.  Comparison of the theoretical pitch angle
curve $i(R)$ with models shows a nice correspondence between the two,
and reveals a strong dependence of the minimum of $i$ on the sound
speed. Already for $\csnd\sim 7 \,\kms$ the nuclear spiral is so
tightly wound that the neighboring shocks become unresolved (see
above).  On the other hand, the nuclear pattern is very open already
at $\csnd \sim 14 \,\kms$.

The nuclear spirals can exist in a fairly narrow range
of gravitational potential shapes and sound speeds. This range is
expected to be typical of the conditions in the central kpc of disk
galaxies. However, we cannot predict what fraction of disks actually
will show this behavior. Partly, this is because a larger
$f_0$ can be compensated by a larger $\csnd$, i.e., more centrally
concentrated potential can correlate with larger sound speeds. A more
detailed parameter search is necessary to decide how common nuclear
spirals should be.

None of the models presented here show nuclear spirals extending to the
center, i.e., they terminate at around $50-100$~pc. At these
distances, the spiral pitch angle is about $30-40^\circ$ and the WKB
theory is not reliable. We note that at such high $i$, the wave packet
will reach the center within less than a quarter of a turn, and the
wavelength is $\sim R$, the distance to the center.

\section{Discussion}

We have presented a way to generate a grand-design spiral pattern in
the central regions of barred galaxies. Such spirals have been
reported recently by Laine \etal. (1999) and Regan \& Mulchaey
(1999). For simplicity, we neglected the gas self-gravity and
investigated the gas response to a background gravitational potential
of a galactic disk with a large-scale stellar bar. The bar pattern
speed and the shape of the gravitational potential were chosen so that
only one ILR exists.  We have found that in such a case, and within a
wide range of parameters, the gas response extends from the corotation
across the ILR into small radii ($50-100$~pc), in a nice agreement 
between the theoretical and numerical morphologies. First, the strength
of the gas response changes substantially at a transition radius, at a
few hundred parsecs from the center. Namely, at larger radii, the gas
displays a pair of strong bar shocks which curve around the ILR.  At
smaller radii, the shock amplitude is substantially reduced after 
first crossing of the bar major axis. Secondly, the pitch angle distribution
shows a spectacular change in behavior at about the same
radius. Outwards of this radius the shock pitch angle increases with
radius, while for the innermost shocks the pitch angle decreases with
radius, in accordance with the density wave theory.
Based on the latter, we argue that the strong outer
shocks perturb the gas inside the ILR, which populates weakly oval
orbits along the bar minor axis. Short waves in the gas propagate
radially inwards and are sheared by the disk rotation, producing weak
shocks.  

We have found that the nuclear spiral pattern depends on two
parameters, i.e., the shape of the precession frequency
$\Omega-\kappa/2$ and the sound speed in the gas. The former is
directly related to the shape of the gravitational potential in the
inner kpc of disk galaxies.  The winding of the nuclear spirals inside
the transition radius increases with a larger radial gradient in the
potential, and hence with the larger central mass concentration. More
centrally concentrated models show much weaker and more tightly wound 
spiral structure which eventually disappears due to the finite numerical
resolution. In realistic disks, this limiting resolution corresponds 
to the finite width of the density wave, given by the mean-free path
of molecular clouds. Damping of density waves results in a featureless 
nuclear gas disk. Decreasing the sound speed of the gas has a comparable 
effect on the nuclear spirals to that of increasing the central mass 
concentration. A higher sound speed creates a more open gas response.

Unfortunately, it is difficult to compare the numerical and theoretical
shapes of nuclear spiral to the observed ones. Measuring the pitch
angle in the $J$-$K$ color index or in unsharp-masking $J$ images 
resulted in a large scatter (Laine, priv. comm). A detailed comparison 
with observations requires more high-quality data and is outside the 
scope of this paper.

Gas gravity should modify the amplitude of the gaseous
response (e.g., Lubow, Cowie \& Balbus 1986), but the basic results of
Sections~2 and 4 are not expected to be grossly violated.  Under
certain conditions, the existence of the so-called $Q$-barrier
(Bertin \& Lin 1989) in a self-gravitating disk can damp both the
short and long waves coming from larger radii to the center, and make
the existence of nuclear spirals unlikely.  In principle, it is
possible to obtain a grand-design spiral pattern in a self-gravitating
nuclear disk by generating it locally. In such a case the pattern
speed of this structure will be gravitationally decoupled from the
large-scale stellar bar and the accompanying large-scale gas response,
as seen in numerical simulations (Heller \& Shlosman 1994). However,
no continuity between the inner and outer spirals is expected in this
case, while at least some of the observations quoted above emphasize
this feature.  Flocculent multi-arm nuclear spirals have been detected
also in disk galaxies which are not known to possess stellar bars.
Those may be formed due to acoustic wave instabilities (Elmegreen
\etal.  1998).

The process discussed here emphasizes the driving (although
indirectly!)  of nuclear structure by a large-scale perturbation
(stellar bar).  This results in the equal pattern speeds for the
nuclear spirals and the large-scale bar, in contrast to all other
mechanisms. Given this, and to the extent that one can neglect the
self-gravitational effects in the wave dispersion relationship
(Eq.~\ref{eq:disp}), detection of nuclear spiral structure limits the
range of parameters, such as the rotation velocity and the sound
speed. For example, in Eq.~\eqref{incl}, observing $\Omega(R)$, $i$ at
some $R$, and {\it assuming} a reasonable sound speed in the gas, will
enable us to put limits on the bar pattern speed, $\OmegaP$, by
Eq.~\eqref{vgrp}.

A number of side effects can accompany the evolution of grand-design
nuclear spirals. We mention only one such effect here. The
characteristic timescale of radial propagation of wave packets across
the nuclear disk (defined here as the area within the transition
radius, $\rt$) is estimated to be about $\sim10^8$~yrs. This timescale is
much shorter than any timescale related to the large-scale galactic
morphology (such as the stellar bar), but it is of the same order of
magnitude as dynamical effects within roughly the central 500~pc.
Those include the random walk of the center of mass of the nuclear
disk with respect to the center of mass of the rest of the galaxy, or
precession of an inclined nuclear disk around the galactic
rotation axis.  In fact, any $m=1$ perturbation in the center will
evolve on this timescale.

It is interesting that the nuclear spiral pattern will respond to
these perturbations slowly enough, so that the shape of the spirals
will not be adjusted instantly. Such a pattern will be able to show
visible distortions. The mere detection of the latter will
tell us about the possible dynamical processes operating in the central
regions of disk galaxies.

\acknowledgments

ACKNOWLEDGEMENTS. We are grateful to Johan Knapen and Seppo Laine for
sharing results with us before publication, to Clayton Heller for the
help with orbital analysis, and to Alar Toomre for enlightening
discussions. We also thank the anonymous referee for suggestions
which helped to improve this manuscript.  This work was supported in
part by NASA grants NAG5-3841 and WKU-522762-98-06, and HST grant
AR-07982.01-96A.

\appendix
% -- appendix must be named 'A', even if there is only one
\section{The frozen potential beyond corotation}

The frozen model potential beyond the corotation consists of three components, 
namely the bulge, the disk, and the halo.
The bulge is described by a Plummer sphere
\eq{
	\Phi_{\rm b}=-{G M_{\rm b} \over \sqrt{r_{\rm b}^2+r^2}}
}
where $r$ is the spherical radius, the scale radius $r_{\rm b}=200\,\pc$, 
and mass $M_{\rm b}=4\times10^9\,\msun$. The
Kuzmin-Toomre disk is given by
\eq{
	\Phi_{\rm d}=-{G M_{\rm d} \over \sqrt{(A+|z|)^2+R^2}}
}
in cylindrical coordinates $(R,z)$, and with $A=3\,\kpc$, and
$M_{\rm d}=2\times10^{10}\,\msun$.
Finally, for the halo we again chose a Plummer sphere with a scale
radius of $r_{\rm h}=9\,\kpc$, and mass $M_{\rm h}=5\times10^{10}\,\msun$.

%-- figures:
\clearpage

\begin{center}
\begin{deluxetable}{lccccc}
\footnotesize
\tablecaption{Spline parameters for the model potentials. \label{tbl-1}}
\tablewidth{0pt}
\tablehead{
\colhead{} & \colhead{Point $i$} & \colhead{$R_i$}   & \colhead{$f(R_i)$}   & \colhead{$f'(R_i)$} 
} 
\startdata
& D & 0 & $f_0\OmegaP$ & 0 \nl
& B & $R_{ILR}$ & $\OmegaP$ & $C_3$ \nl
& C & $R_{CR}$ & $E_1$ & $E_2$ \nl
\enddata
\end{deluxetable}
\end{center}

\end{document}